# Determination of the critical current density in the *d-wave* superconductor YBCO under applied magnetic fields by nodal tunneling

Roy Beck, Guy Leibovitch, Alexander Milner, Alexander Gerber and Guy Deutscher

School of Physics and Astronomy, Raymond and Beverly Sackler faculty of Exact Sciences, Tel-Aviv University, 69978 Tel-Aviv, Israel

PACS: 74.25.Sv, **74.50.+r,** 74.72.Bk

**Abstract**

We have studied nodal tunneling into $YBa_2Cu_3O_{7-x}$ (YBCO) films under magnetic fields. The films' orientation was such that the $CuO_2$ planes were perpendicular to the surface with the a and b axis at $45^0$ form the normal. The magnetic field was applied parallel to the surface and perpendicular to the $CuO_2$ planes. The Zero Bias Conductance Peak (ZBCP) characteristic of nodal tunneling splits under the effect of surface currents produced by the applied fields. Measuring this splitting under different field conditions, zero field cooled and field cooled, reveals that these currents have different origins. By comparing the field cooled ZBCP splitting to that taken in decreasing fields we deduce a value of the Bean critical current superfluid velocity, and calculate a Bean critical current density of up to $3 \cdot 10^7$ $A/cm^2$ at low temperatures. This tunneling method for the determination of critical currents under magnetic fields has serious advantages over the conventional one, as it avoids having to make high current contacts to the sample.

As shown by de Gennes and Saint James (1), finite energy bound states are formed in a normal metal film (N) in contact with a conventional (*s-wave* symmetry) superconductor (S). These states have energies smaller than the gap in S. They correspond to Saint James cycles (2), in which an electron is converted at the N/S interface into a reflected hole (3), which undergoes a specular reflection at the outer surface of N before being Andreev reflected into an electron at the interface, and finally specularly reflected, thus completing the cycle. Similar Saint James cycles occur when S is a *d-wave* superconductor, but with one major difference. As shown by Hu (4), if the interface is perpendicular to a nodal direction, zero energy bound states are formed because of the phase difference of $\pi$ between successive Andreev reflections. These zero energy states persist in the limit where the normal metal thickness is zero. They produce the ZBCP in the nodal tunneling conductance (5).

As shown by Fogelstrom *et. al.* (6), surface currents flowing in the $CuO_2$ planes break time reversal symmetry and induce a Doppler shift (DS) of the zero energy states, resulting in a split of the ZBCP. Such currents can be Meissner's origin generated by a magnetic field applied parallel to the surface (after zero field cooling of the sample), as in Lesueur *et. al.* (7) and Covington *et. al.*(8). Meissner currents increase at first linearly with field, and eventually up to fields of the order of the thermodynamical field $H_c$ if the Bean Livingston barrier (9) is effective, and so should the splitting given by $\delta_\uparrow(H) = v_S p_F \sin\Theta$, where $v_S$ is the superfluid velocity of the Bean Livingston currents, $p_F$ the Fermi momentum of tunneling particles and $\Theta$ the width of the tunneling cone. This linear behavior has been observed experimentally (7,8,10). Moreover, as shown by Krupke and Deutscher (10) and by Aprili et al. (11), splitting does not occur when the field is applied parallel to the $CuO_2$ planes, which is a strong indication that the ZBCP is indeed due to the *d-wave* symmetry.

The correlation between surface currents and splitting of the ZBCP having been previously established experimentally under conditions of increasing fields, we show here how it is possible from splitting measurements to determine surface current densities under different conditions. Specifically, the difference between splitting values measured under field cooled (FC) and decreasing field conditions gives access to the Bean critical current density. We find that at 4.2K it does not vary significantly in YBCO films up to fields of 15T, remaining equal to about $3 \cdot 10^7$ a/cm$^2$.

Experimental

We have formed tunneling junctions by pressing an indium wire on a fresh film of off-axis sputtered $Y_1Ba_2Cu_3O_{7-x}$ as described elsewhere [12]. Films thickness was 1600Å and 3200Å, with a $T_c$ of 89±2 K. The films were in-plane oriented, with the [0,0,1] axis oriented parallel to one of the film's edges, and the [1,-1,0] axis oriented parallel to the other one. The crystallographic direction normal to the surface, in which the tunnel junction was directed, was [1,1,0], as verified by x-ray diffraction. Scanning electron microscopy and atomic force microscopy of the films surface revealed well oriented crystallites, with a surface roughness of a few nm. We have measured I/V curves using DC a current supply and a digital voltmeter and calculated the differential conductance (i.e. dI/dV vs. V) numerically without any data manipulation such as averaging of the readout data. Junctions' quality was verified by observing the Indium superconducting gap feature at zero magnetic field at $T<T_c$ of the Indium (3.4 K). The magnetic field was applied parallel to the films surface and along the c-axis direction, perpendicular to the $CuO_2$ planes.

Results and discussion

Fig.1 shows splitting values of the ZBCP measured under FC conditions, together with values taken in zero FC in increasing and decreasing fields for a 1,600Å thick sample at 0.3K. FC values are smaller than those taken in increasing fields, and larger than those taken in decreasing ones, being closer to the latter. This order can be understood if we assume the existence of a field induced imaginary component of the order parameter, possibly having the $id_{xy}$ symmetry as argued by Beck *et al.*(12). Such a component induces a splitting equal to its amplitude. In increasing fields, the splitting is enhanced by the additional contribution of the Doppler shift generated by the Bean Livingston (BL) currents as discussed above. In addition, there may also be a contribution of the Bean critical state currents due to bulk pinning, usually weaker than that of the BL currents and flowing in the same direction. Under FC conditions, there are neither BL nor Bean currents, the dominant contribution to the splitting is that of the imaginary component. Under decreasing fields, there are no BL currents (there is no BL barrier against flux exit), but there will be reversed Bean currents. The difference between the FC and the decreasing field data is then due to the Bean surface currents.

We show in Fig.2 the FC data as a function of $H^{1/2}$. It is similar to that of data taken in decreasing fields, also shown. Both sets of data can be fitted to straight lines having a slope of 1.1 mV/T$^{1/2}$, in agreement with Beck *et al.*(12). The origin of this field dependence has been discussed elsewhere (12). Briefly, it has been attributed to a field induced $id_{xy}$ component of the order parameter, as originally proposed by Laughlin (13). Actually, under FC conditions, an additional contribution to comes from the finite equilibrium screening currents corresponding to the bulk magnetization. These currents are of the order of the difference between the induction and the applied field, divided by the distance from the surface to the first vortices, itself of the order of the intervortex distance (14). In decreasing fields, the Bean critical state currents run in a direction opposite to that of the equilibrium currents, thus reducing the splitting. What we wish to exploit here is the difference between the two sets of data. Since they fit parallel lines, the Bean critical current must be field independent in the range of fields investigated. The constant difference between the two lines which is of about 0.6 mV for that 1600Å thick sample. We can use this value to calculate the velocity of the Bean currents using the expression given by Fogelstrom *et.al.* (6):

(1) $$\delta_{FC} - \delta_\downarrow = v_s(B) \cdot p_F \sin\Theta$$

Where $\delta_{FC}$ and $\delta_\downarrow$ are the splitting in FC and in decreasing field respectively, and $v_S(B)$ is now the superfluid velocity *of the Bean critical state currents*, rather than that of the BL currents as in ref.6. Taking the wave vector of the Fermi surface to be $k_F = 1 \cdot 10^8$ cm$^{-1}$, and estimating the tunneling cone to be $\Theta = 20°$, we calculate $v_S = 2.7 \cdot 10^4$ cm/s. Then, from: $j_c = nev_S$, with a superfluid density, $n = 5 \cdot 10^{21}$/cm$^3$, we obtain for the critical current density $j_c = 2.2 \cdot 10^7$ A/cm$^2$ at 0.3K. For a somewhat thicker sample (3200Å), the difference between the FC and decreasing fields data is of 0.9 meV and we obtain $j_c = 3 \cdot 10^7$ A/cm$^2$ at 4.2K. These values are in good agreement with direct measurements (15). It turns out that the Bean currents are of the same order as that of the calculated equilibrium screening currents under field cooled conditions (but of course of opposite direction). This may explain why the straight line running through the decreasing field splitting data in Fig. 2 extrapolates closer to zero at zero field than the FC data does.

Conclusions

We have measured the nodal tunneling conductance into YBCO films under magnetic fields applied parallel to the surface and perpendicular to the $CuO_2$ planes. We have shown that the Bean critical current density can be obtained by taking the difference in the splitting of the Zero Bias Conductance Peak, in a given magnetic field, between FC and decreasing field data. We have found this critical current to be field independent and equal to about $3 \cdot 10^7$ A/cm$^2$ up to 16T at low temperatures. This novel method avoids having to pattern the sample, thus removing the difficulties associated with possible heating effects under strong injected currents.

Acknowledgements

We are indebted to Roman Mintz and Vladimir Kogan for an enlightening discussion on the field dependence of screening currents under field cooled conditions. This work was supported by the Heinrich Hertz Minerva Center for High Temperature Superconductivity, by the Israel Science Foundation and by the Oren Family Chair for Experimental Solid State Physics.

Figures :

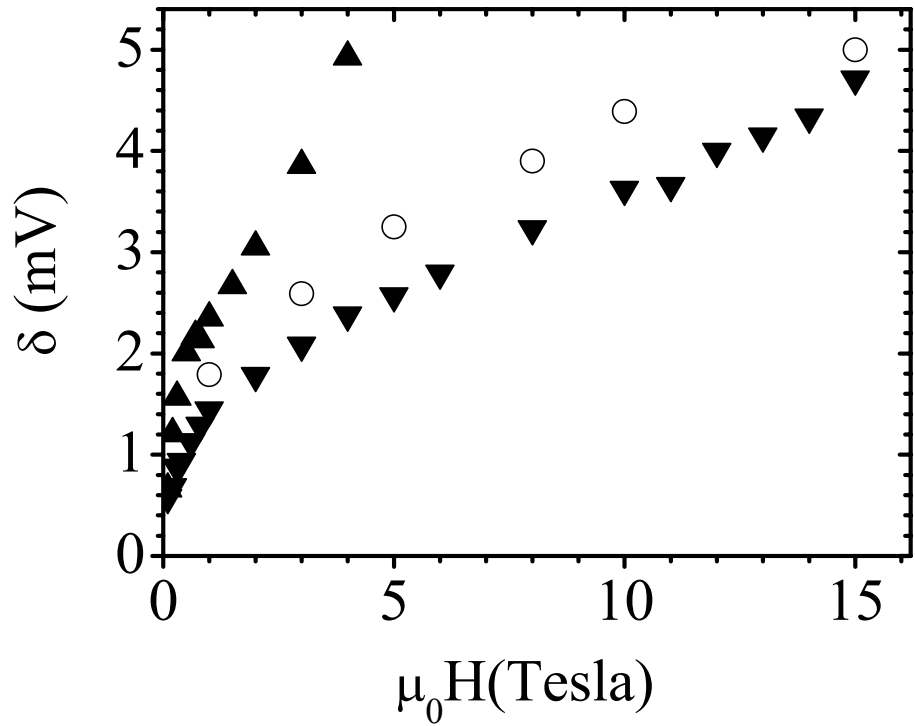

Figure 1: ZBCP splitting, δ, as a function of magnetic field applied parallel to the film surface and perpendicular to the $CuO_2$ plains at 0.3K. Circles indicate measured splitting in field cooled conditions, and triangular for zero field cooling and increasing (pointing up) and decreasing (pointing down) fields in the range of 0 to16T.

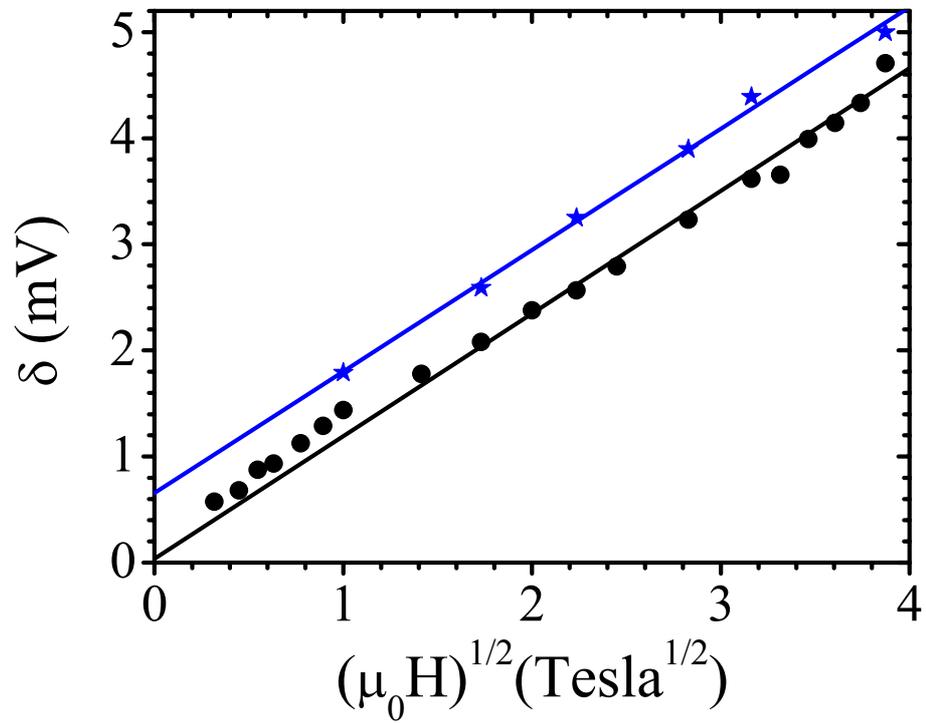

Figure 2: ZBCP splitting, $\delta$, as a function of the square root of the applied magnetic field, for field cooled conditions (stars) and decreasing field conditions (circles). Linear fits to the data (lines) have slopes $1.14\pm0.03$ mV/T$^{1/2}$ and $1.16\pm0.02$ mV/T$^{1/2}$ and intersections of $0.66\pm0.09$ mV and $0.03\pm0.02$ mV for field cooled and decreasing field respectively.